# Tribological Analysis of Ventral Scale Structure in a Python Regius in Relation to Laser Textured Surfaces

H. A. Abdel-Aal[*]
Arts et Métiers ParisTech, LMPF-EA4106,
Rue Saint Dominique BP 508,
51006, Châlons-en-Champagne, France
[*]*Hisham.abdel-aal@ensam.eu*

M. El Mansori
Ecole Nationale Supérieure d'Arts et Métiers,
151 Boulevard de l'Hôpital, 75013 PARIS, France

**ABSTRACT**

Laser Texturing is one of the leading technologies applied to modify surface topography. To date, however, a standardized procedure to generate deterministic textures is virtually non-existent. In nature, especially in squamata, there are many examples of deterministic structured textures that allow species to control friction and condition their tribological response for efficient function. In this work, we draw a comparison between industrial surfaces and reptilian surfaces. We chose the python regius species as a bio-analogue with a deterministic surface. We first study the structural make up of the ventral scales of the snake (both construction and metrology). We further compare the metrological features of the ventral scales to experimentally recommended performance indicators of industrial surfaces extracted from open literature. The results indicate the feasibility of engineering a Laser Textured Surface based on the reptilian ornamentation constructs. It is shown that the metrological features, key to efficient function of a rubbing deterministic surface, are already optimized in the reptile. We further show that optimization in reptilian surfaces is based on synchronizing surface form, textures and aspects to condition the frictional response. Mimicking reptilian surfaces, we argue, may form a design methodology potentially capable of generating advanced deterministic surface constructs capable of efficient tribological function.

**Nomenclature**

| | |
|---|---|
| hp | Dimple height |
| hs | Denticulation (fibril) height |
| Ra | Average Roughness height |

*Symbols*

| | |
|---|---|
| $\lambda$ | Fibril row intra spacing (μm) |
| $\lambda p$ | Intra Spacing on Laser Textured Surface |
| $\lambda s$ | Intra Spacing on Snake Skin |
| $\Phi$ | Dimple Base diameter |
| $\Phi_s$ | Dimple Base diameter for snake skin |

*Abbreviations*

| | |
|---|---|
| AE-PE | Anterior Posterior Axis |
| BDP-S | Bottom Dead Position Snake |
| COF | Coefficient of Friction |
| FIEL | Friction-Induced Energy Losses |
| FAR | Fibril Aspect Ratio |
| LL | Left lateral direction |
| LTS | Laser Textured Surface |
| MP-S | Meddle position-Snake |
| RL | Right lateral direction |
| VSAR | Ventral Scale Aspect Ratio |
| TAR | Total Area ratio |
| TDP-S | Top Dead Position Snake |
| SAR | Surface Aspect ratio |
| DSR | Dimple Slenderness Ratio |
| LCR | Length to Circumference Ratio |

**Introduction**

Friction-Induced Energy Losses, FIEL, of a rubbing system has two contributions. The first is a result of friction between the micro-topography at the interface between the contacting bodies. The second is a consequence of the friction between the lubricants, if present, with the interface. The magnitude of the second component increases upon using a lubricant with high viscosity (which is necessary to support high frictional loads). Reduction of the frictional tractions allows using lubricants of lower viscosities and thereby it reduces the losses due to lubricant friction. Therefore, currently, many efforts address the possibility of engineering topographies in order to improve the quality of surface-interaction in rubbing assemblies. Successful engineering of surface topography, therefore, leads to reduction in the overall FIEL.

Ideally, the target is to engineer surfaces that yield predetermined rubbing response, and are, in the same time, capable of self-adapting such response in accordance with changes in sliding conditions. Such surfaces, termed as "deterministic surfaces" comprise artificial textures embossed on the rubbing interface. The texture building block is a micron-sized 3-Diensional geometrical shape (cone, hemisphere, rounded apex, chevron etc.,) which repeats as an array over the desired area of the surface.

There are several techniques to produce these textures (e.g., multistep honing, helical-slide honing, controlled thin layer deposition [1-4], and laser texturing [5-9] which is the most advanced and is considered by many as a promising enabling technology [10-13].

Although available since the seventies of the twentieth century application of Laser Texturing to frictional surfaces however, began early this century when it was initially applied to mechanical seals [5-6] then to piston rings and cylinder bores [7,8]. The process involves creation of an array of micro-dimples, either positive (protruding above) or negative (carved into) the target surface using a material ablation process with a pulsating laser beam. Theoretical analysis identified several dimensional groups that influence the tribological performance of a textured surface [15-26]. To date, however, there is no agreement on the optimal values a particular surface should acquire. More importantly, a well-defined methodology for the generation of textures for optimized surface designs is virtually non-existent because of the absence of a holistic surface-design methodology that merges function, form and topography to achieve lean performance. While Surface Design Optimization, in essence, has not matured, as of yet, within the realm of human engineering it is advanced in natural designs especially within the scaled reptiles (squamata).

Squamate Reptiles present diverse examples where surface structure, texturing, and modifications through submicron and nano-scale features, achieve frictional regulation manifested in: reduction of adhesion [27], abrasion resistance [28], and frictional anisotropy [29].

Squamata comprises two large clades: Iguania and Scleroglossa. The later comprises 6,000 known species, 3100 of which are referred-to as "lizards," and the remaining 2,900 species as "snakes" [30]. Snakes are found almost everywhere on earth. Their diverse habitat presents a broad range of tribological environments. This requires customized response that manifests itself in functional practices and surface design features. This, potentially, can inspire deterministic solutions for many technical problems. Many authors studied appearance and structure of snakeskin in relation to friction. Results point at the relation of surface topography in snakes to tribological performance [29-41].

The motion of a snake is a delicate balance between the propulsive forces generated by the muscles and the friction tractions due to contact with the substratum. In some cases, the snake makes use of friction to generate thrust. However, for economy of effort, the COF needs to be minimized (especially in rectilinear locomotion) since friction opposes motion. As such, a self-regulating mechanism to control frictional tractions should exist in the snake. The texture of the surface (i.e., the micron sized fibril structures or denticulations) are a major component of such a mechanism. The geometry and topology of the fibril structures allow the snake to modify the frictional profiles in response to changes in contact situations. The presence of the fibrils contributes to the dynamic control of the real area of contact between the skin and the substratum upon sliding [42]. The function of the denticulations (micron-sized fibrils) in this sense is similar to that of the deterministic textures used to regulate friction. Such a similarity raises a curious question. Namely, if we consider the denticulations present on the ventral side of a snakeskin as deterministic textures, would their descriptive parameters coincide with, or at least match, the range of values recommended by researchers based on laser texturing? Furthermore, in case of the validity of such a preposition, can the texturing of a snake inspire a new paradigm in deterministic surface texturing?

There are several similarities between a LTS and the ventral side of a snake, both in metrology and topology. The main similarity is the dependence of the metrological features on the scale of observation. That is both of the snakeskin and the LTS belong to the so-called multi-scale surfaces [40]. In addition, the basic building block in each case is a textural element that repeats in an array. The snake's ventral side is composed of identical micron-sized fibrils (denticulations) distributed over the skin area in a particular pattern. Spacing, length, orientation and shape of denticulation are, in general, common to a particular family of snakes. LTS, on the other hand, by their very definition, comprise an individual textural building block (cone, dimple, chevron etc.,) that also forms an array on the surface. Therefore, both types of surfaces share a common constructal origin. A snake, however, has to be self-sustaining over a wide spectrum of sliding terrains and conditions. The ventral texturing, therefore, has to be efficient over the spectrum of contact conditions that the species experiences. Such a feature offers an advantage to the natural surface over an engineered LTS where the dimples satisfy efficient performance over a narrow domain within possible contact conditions. The question, therefore, becomes how to optimize LTS in order to extend efficient performance, similar to that of a snake. In other words, how to capture the essence of texturing within the ventral side of a snakeskin and then incorporate it into a manmade surface to enhance tribological performance.

The answer to this question is rather complex due to the many factors involved. A simple answer is to mimic, or replicate, the textural designs present on the reptile in a technological surface. However, the success of this approach depends on the existence of a one-to-one match between the function and perhaps, the material, of the target surface and that of a reptile. This match will limit the replication process to polymeric surfaces (because the skin is elastomeric).

However, bio-inspired surface texturing has a broader goal than merely replication. In essence, it seeks to extend the potential tribological benefits of reptilian surfaces to the domain of man-engineered surfaces. Transfer of design benefits requires developing multi-aspect compatibility metrics in order to evaluate the feasibility of the inspiration process itself (i.e., basing the design of a technical surface on that of a biological analogue). Two features are important in this context: the quality of frictional performance of the snake and matching of the deterministic surface parameters of the biological target to those recommended for the technological surface.

Frictional performance of snakes was a subject of several investigations that confirmed the unique features of the ventral skin of snakes and their optimized tribological response with respect to energy losses and resistance to abrasion and wear [36, 42-44]. Other results [43, 45, 46] attribute optimization of tribological function to the geometrical patterns and metrology of the ventral micro-texture. These results, while promising, are to be complimented by assessing the deterministic features of the ventral textures along the same guidelines followed in examining LTS (and manmade surfaces in general). This study, therefore, presents a comparative study of the deterministic surface parameters of the ventral texturing in a snake (Python regius) and those parameters recommended from LTS. The choice of the python species for this study aims enhances the reliability of the comparison results due to the similarities between the mode of motion of this species and the kinematics of many rubbing surfaces.

Pythons manifest some of the heaviest constrictor snakes. Their length and weight limits their locomotion to the rectilinear mode [47]. Due to this limitation, the snake depends on continuous frictional adaptation for propulsion. It also depends mainly on the ventral side for propulsion, which implies that the textural denticulations mostly encounter linear friction. This is similar to the mode of friction dominant within many rubbing contacts where linear relative motion, between complying surfaces, takes place. Moreover, the shape of the tips of the denticulations resembles a dimple (spherically capped asperity). Such a shape is the most studied within LTS literature. This work is organized in three parts. The first part details the general appearance and structure of the ventral skin of the Python. The emphasis is on the dimensional metrology of the micron-sized denticulation structures on the ventral side. Following this part, we present the essential deterministic metrology of the texture. The third part of the paper presents the detailed comparison between the parameters of the python and those recommended for LTS. In this part we further, evaluate the feasibility of the idea of building a tribological surface based on Python texturing and we identify several design lessons deduced from the texturing of the snake.

## 2. Laser Textured Surfaces
## 2.1 Description of surface Features

Surfaces entail features of many different scales. Any surface engineered to meet a predetermined functional requirement, such as enhanced lubrication or reduced frictional response, includes surface features of different sizes and distribution. The distribution of the surface features, along with the type, will also determine the method of describing the performance metrics. Traditional surfaces (surfaces that result from conventional manufacturing processes such as grinding, turning etc.,) the organization of the surface features (height, intervals, size etc.,) is stochastic in essence. Description of surface metrics in such a case stems from signal processing techniques (e.g., spectral analysis, auto correlation functions, root mean square height etc.). The reason being that on a fundamental level, the stochastic elements resulting from the surface generation process dominate the texture. When, on the other hand, a surface contains dominant deterministic patterns (such as tessellations, axially symmetrical patterns, rotationally symmetric patterns) a stochastic description of the surface features becomes meaningless [48].

The conventional approach for describing the compositional features of a surface invokes statistical techniques. Consequently, the distinctive property of any texture descriptor, for a given stochastic surface, is that this descriptor is a statistical variable (i.e., mean, maximum, root mean square, Skew etc.). The value of a characteristic texture feature in this frame is the expected value of the relevant statistical variable (s) or, in some cases, is the relationship between values of two statistical variables. By invoking statistics, one can account for individual variations between all textural features of the same kind present in the studied surface. For example, one can describe the roughness of the surface by accounting for the variation in heights of all the asperities present on the surface. The roughness feature in this case, Ra, will be the statistical average of the heights of all summits present on the surface. In the case of deterministic surfaces, however, the quasi-invariance in the size of surface features renders statistical analysis meaningless. This is because in the limit, the value of the statistical variables, conventionally used to express surface features, will converge to the size of the described feature. In other words, the statistical description of surface features of a deterministic surface will converge to the scales and dimensions of the building elements of the texture (which are deterministic to start with).

## 2.2 Descriptive metrics for dimpled surfaces

For the simplest shape, the dimple, the height of the dimple (hp), dimple base diameter ($\Phi_p$), and the center line-to-centerline spacing between dimples $\lambda$ describe the texture of the surface (figure 1).

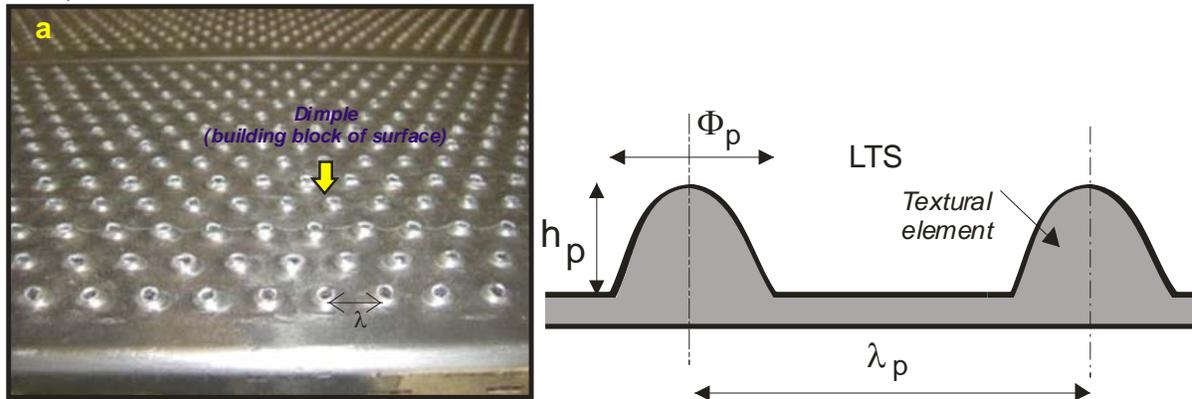

*Figure 1 Definition of the primitive geometrical attributes for Laser Textured Surfaces (a) a sample laser textured (dimpled) surface The height of the dimple (h), Dimple base diameter ($\Phi$), and the center line-to-centerline spacing between dimples $\lambda$.*

Many authors, [5-8, 49-52] use these descriptors to identify three ratios as key performance indicators for the functional quality of LTS. These are the Total Area Ratio (TAR), the Dimple Slenderness Ratio (DSR), and the Surface Aspect Ratio (SAR) (see table 1).

Table 1 Definition and calculation formulae of the main parametric relations used to describe the deterministic features of a dimpled LST (all symbols defined in figure 1)

| Parameter | Definition | Formula |
| --- | --- | --- |
| Total Area Ratio | Total area of the surface occupied by the texturing element to the total area of the surface | $\frac{\text{Total area of dimples}}{\text{total surface area}}$ |
| Dimple Slenderness Ratio (DSR) | Ratio of the height to the diameter of the dimple | $DSR = \frac{h_p}{\Phi_p}$ |
| Surface Aspect Ratio (SAR) | Ratio of the centerline-to-centerline spacing between dimples to the height of the dimple | $SAR = \frac{\lambda_p + \Phi_p}{h_p}$ |

## 3. The Python species

*Python regius* is a non-venomous species native to West Africa. The build of the reptile, figure 2-a, is non-uniform (i.e., the ratio of the body length to the diameter of the body varies along the AE-PE axis. The head-neck region and the tail region are relatively thinner than the main portion of the body (trunk). The trunk comprises the principle load bearing region of the body (i.e., it is where most of the generation of frictional tractions takes place). Table-2 gives a summary of the taxonomic rank of the reptile as well as its' major characteristics.

Table-2 Summary of the taxonomy of the Python regius species and main geometrical features of the reptile

| | Python regius |
|---|---|
| **Taxonomy** | |
| Family | Pythonidae |
| Subfamily | Python |
| Genus | P. regius |
| **Body Geometry and dimensions** | |
| Maximum Length (cm) | 150 |
| Number of ventral scales | 208 |
| Ratio of length to maximum diameter | 10.2 |
| Mass (Kg) | 1.3 |

The ventral side of the reptile comprises hexagonal scales that are elongated along the lateral axis of the reptile (see figure 2-a). The areas of individual scales vary along the AE-PE axis of the reptile. Individual ventral scales, although hexagonal in shape, are not straight edged. Rather their boundaries are arcs and not straight lines (see figure 2-b) and the arcs are curved toward the posterior end. The cross section of the body of the reptile parallel to the Dorso-Ventral axis is almost parabolic with the ventral side protruding outwards. This curved segment, comprising the ventral scale, curve A-A in figure 1-c, supports contact tractions and the weight of the reptile. The length of the curve A-A varies by location and therefore the ratio of the length to the circumference of the reptile varies along the AE-PE axis.

The ventral side comprises two distinct regions scales and hinges. The scales are relatively rigid (almost like a membrane), whereas the hinge region is flexible. Scales contain micro fibril structures (also known as denticulations).

The hinge region, on the other hand, contains a series of pores. Figure 3, depicts the composition of the scale region (figures 3-b, d, and e) and that of the hinge region (figures 3-a, c, and e) at different magnifications. The denticulations form wave-like rows separated by a distance, λ. The fibrils within the scale region point in the general direction of the posterior end (head-to-tail). Detailed description of the structural features of the ventral scales is given elsewhere *[41, 42]*. Here, however, we discuss those features pertaining to the current work.

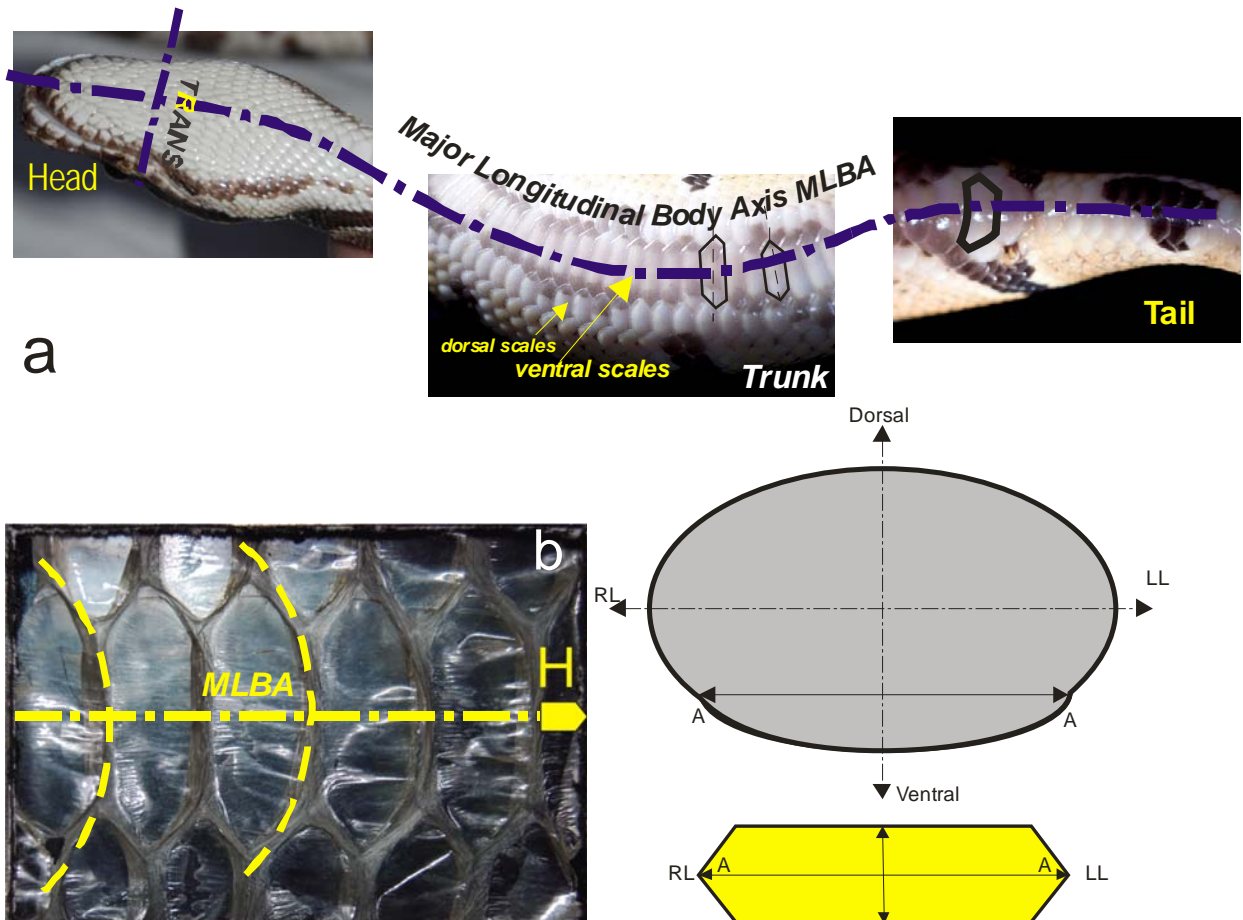

*Figure 2: The ventral side of the snake: a- definition of the axis chosen to describe relative position of the ventral scales on the snake body, b- magnification of the ventral scales located within the mid-section of the reptile. The letter H denotes orientation of the head, c-schematic of a cross section of the body of the reptile along the dorso-ventral axis, and a plan view of a generalized ventral scale (AA is the lateral axis of the scale and BB is the longitudinal axis of the scale).*

The stocky build of the python regius species manifests distinctive change in the ratio of the snout-to-vent length to the circumference. In this work, we used this physical feature to determine the boundaries of the load bearing volumes on the body of the reptile. To this end, we define three regions. The first region, TDP-S, represents the top boundary of the load-bearing volume within the body. The highest LCR characterizes this region. The second, region, MP-S, represents the medium portion of the reptile body, and is the bulkiest of all regions. The third region, BDP-S, represents the lower boundary of the load-bearing volume of the body. Similar to the TDP-S this region also has a constricted circumference and a high LCR. Table 3 presents a summary of the location of the defined region along the AE-PE-axis and the defining geometrical rations.

The geometry of the micron-sized fibrils within the three regions is similar. However, the spacing and density of the fibrils (and thereby the ratio of area occupied by fibrils to the rest of the area of a scale) will also differ in each region. Figure 4, presents two sets of SEM pictures depicting ventral scales located within three regions on the ventral side denoted as TDP-S, MP-S, and BDP-S. The choice of these regions follows from the general build of the reptile. Table 3 presents a summary of the location of these regions on the AE-PE axis along with the LCR, which varies considerably at the beginning of each region. Variation in the LCR implies an analogous variation in bearing the capacity of friction-induced loads of each region.

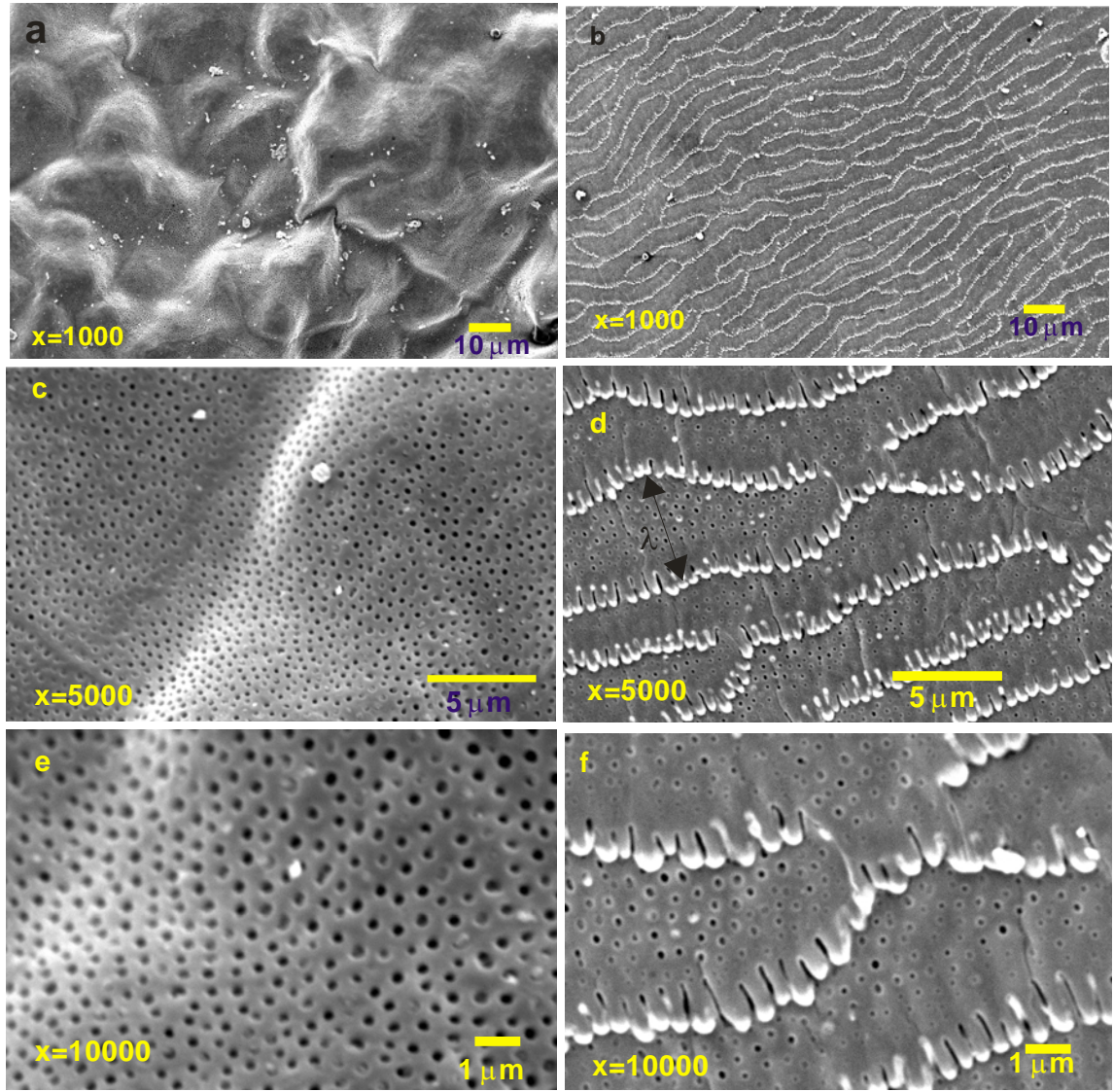

*Figure 3 Details of the structure of the ventral scales of the Python regius species (a, c, and e) different magnification of the hinge region, (b, d, and f) details of the "scale" region.*

Table 3 Summary of non-dimensional location and geometrical features of the body regions shown in figure 4 on the AE-PE

| X/L | Zone | Lat. Diag. (cm) | Circumference (cm) | Ratio of length to circumference |
|---|---|---|---|---|
| 0.01 | TDP-S | 1.5 | 6.75 | 22.25 |
| 0.67 | MP-S | 2.1 | 10.5 | 14.2857 |
| 0.325 | MP-S | 1.95 | 9.75 | 15.385 |
| 0.775 | BDP-S | 1.6 | 7.845 | 19.125 |

## 4. Analogy between LTS and Snake skin

Snakes use at least five unique modes of terrestrial locomotion (lateral undulation, side winding, concertina locomotion, rectilinear locomotion, and slide pushing). Each of these modes has unique energetic, as well as mechanical, requirements. Although there are distinct kinematic differences between the individual modes of motion, they all share their origin in muscle activity. Transfer of motion between the active muscle groups and the contacting substrate will thus depend on generation of sufficient tractions. The skin handles generation of tractions and accommodation of motion. The skin of a snake while transferring locomotion tractions also has to accommodate the energy consumed in resisting the motion.

Muscular activity for locomotion comprises sequential waves of contraction and relaxation of appropriate muscle groups. The number, type, and sequence of muscular groups responsible for the initiation, and sustainment, of motion, and thus employed in propulsion vary according to the particular mode of motion. This implies that contact stiffness in dynamic as well as in static friction constantly varies. Generation of tractions for motion also depends on the habitat and the surrounding environment. This, in turn, affects the effort invested in initiation of motion and thereby affects the function of the different parts of the skin.

In figure 4, left hand side pictures reveal fibril structures whereas, right hand side images provide details of the fibrils. The fibrils protrude over the background of the ventral scale and are arranged in waves rather than in orthogonal arrays (perpendicular rows and columns) as often practiced in man-manufactured surfaces. The spacing between fibrils appears to be non-uniform and the size of the individual fibrils seems to vary by location.

Due to their relatively heavy weight, pythons use rectilinear locomotion (movement in a straight line). In this mode, the reptile slightly lifts the ventral scales from the ground, pulls them forward, and then downward and backward. However, because the scales "stick" against the ground, the body moves forward over them. Once the body has moved far enough forward to stretch the scales, the cycle repeats. This cycle occurs simultaneously at several points along the body.

In rectilinear motion, the main force required is that to overcome external friction and for acceleration of the various regions of the body. An equal and opposite static reaction balances the forward propagation force under the stagnant segments fixed to the ground. Furthermore, the highly developed ventral coetaneous musculature, that generates a peristaltic wave along the snake body, facilitates locomotion. These contraction waves include modulation of both the area and pressure of contact between the skin and the ground [47] through interaction between the ventral skin and the substrate. In particular, the geometrical asymmetry of fibril tips induces precise adaptation of the frictional response [37-39, 41, 42, 53, 54].

AFM imaging of the fibrils, shown in figure 5 (a and b), reveals the elevation of the tips above the overall ventral cell plateau. In this sense, the tips resemble positive dimples raised above the general plane of a textured surface (similar to the micro-dimples created in a laser texturing process). A distinct difference however is the arrangement of the texturing in each case. For the shed skin, the distribution of the fibril tips (dimples) on the scale manifests wave arrangement (as opposed to the orthogonal array arrangement observed in manmade surfaces). The function of texturing is to modify the frictional behavior of the surface upon sliding. In this sense, both manmade and snake surface share the origin of tribological response modifiers: texturing. This observation provides the essence of the comparison between the topological features of the skin of a Python and an LTS.

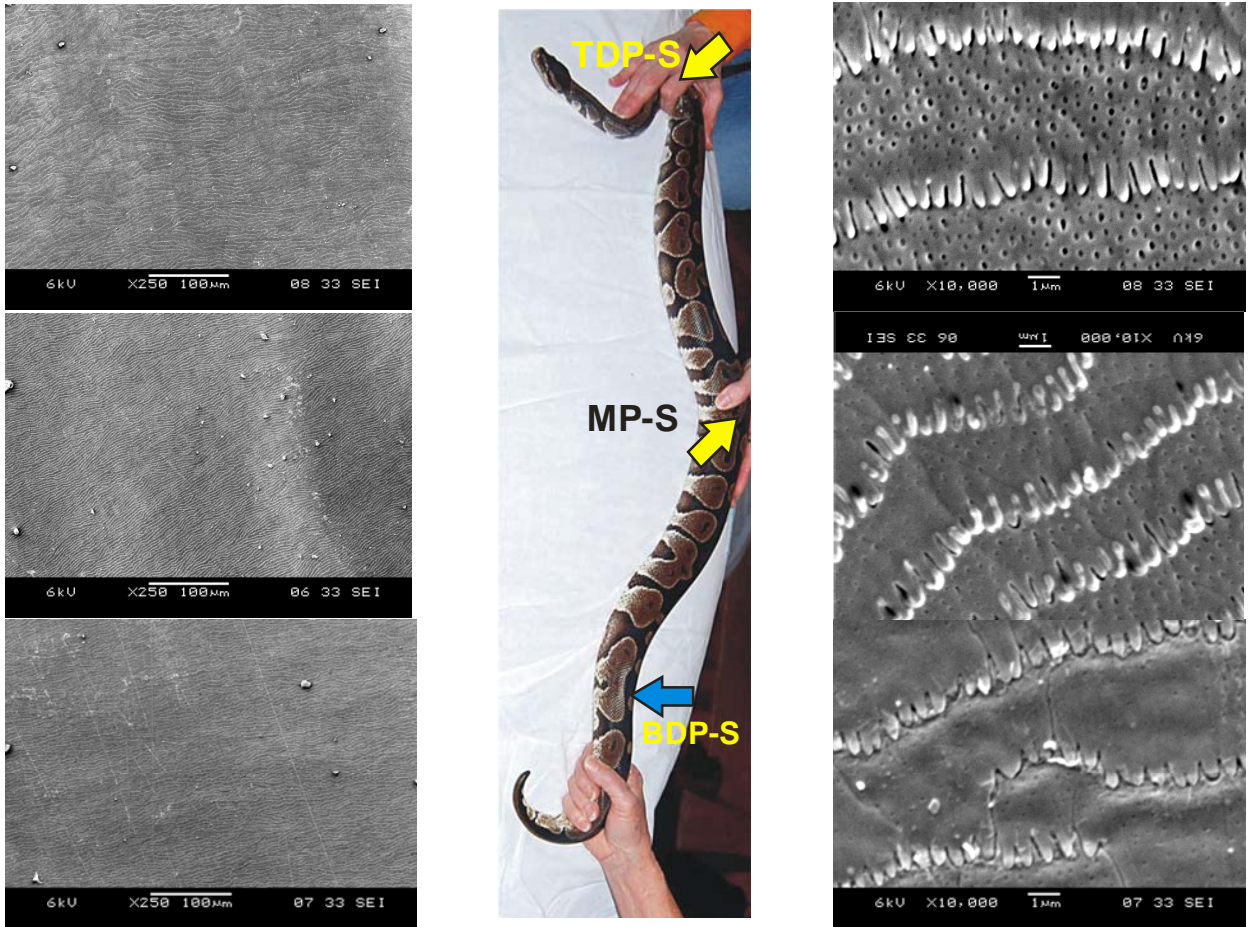

*Figure 4. SEM images depicting structural details of the ventral scales at three locations, TDP-S, MP-S, and BDP-S respectively. Images on the left hand side are taken at a magnification of X=250 (length marker 100 μm) and those on the right hand side are taken at a magnification X=10,000 (length marker 1μm). Note the arrangement of the fibrils in wave like rows, and the non-uniform distribution of the fibril dimensions.*

To compare the texture of the snakeskin to that of the LTS, there is a need to define texture descriptors for reptilian skin along the same lines used for LTS. The direct approach is to redefine the LTS descriptors in terms of the surface features present on the snakeskin. To this end, we use the subscript (s) to denote surface texture features present on the snakeskin, where as we use the subscript (p) to denote descriptor of LTS. Therefore, for the snakeskin the DSR is defined as the ratio of the fibril height ($h_s$) to the base diameter of the fibril tip ($\Phi_s$). The SAR meanwhile is redefined as the distance between two consecutive fibril waves ($\lambda_s$) averaged over the entire body and the protrusion distance. Finally, the TAR for the snake is redefined as the total area of the fibrils to the total surface area. Table 4 presents a summary comparison of the formulas used in computing the additional surface parameters for an LTS and a snake.

The total fibril height of the snake, $h_s$, manifests the protrusion of the fibril tip over the cell plateau. This is approximately equivalent to the average roughness of the surface, Ra, along the orientation of the fibrils (i.e., along the AE-PE-axis).

According to the preceding, to compare the performance indicators of the shed skin to those of dimpled LTS one needs to calculate the four parametric ratios listed in table 4 for each surface.

Note that the descriptors of an LTS depend on the primitive geometry of the surface, those for a shed skin, however, are functions of metrological parameters (dimensional and textural). As such, for the shed skin, characterization of the essential metrological parameters should precede the calculation of the performance indicator ratios. Once the necessary parameters are determined, calculation of the four ratios of table 4 may take place.

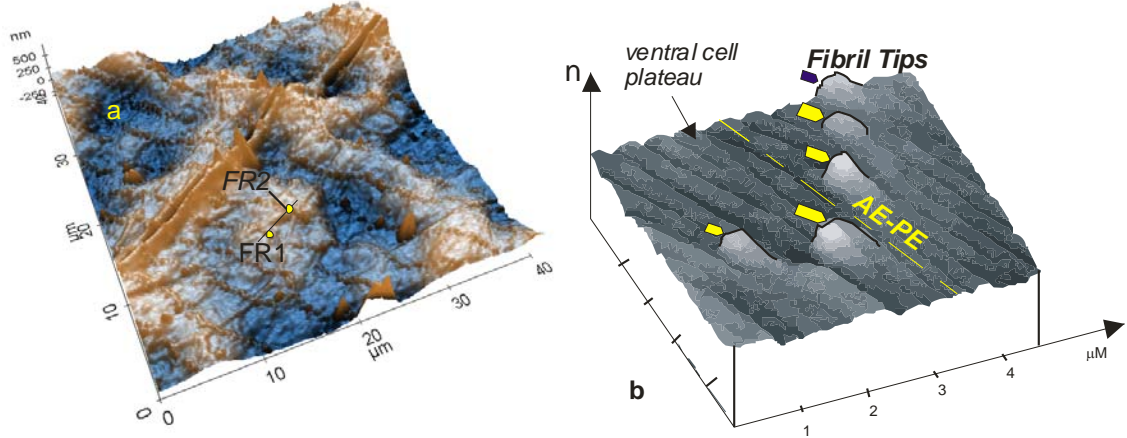

*Figure 5: Three dimensional AFM-images of the fibril structures within the ventral scales of the Python regius, a- general scan of a 45 μm x 45 μm area of a ventral scale. Yellow dots mark two fibril rows (FR1 and FR2). Note the protrusion of the fibrils above the general plane of the scale, b- AFM-image of a 5 μm x 5 μm area of the ventral scale. Note the orientation of the fibrils along the AE-PE Axis*

Table 4: Summary of formulas used to calculate parameters used in comparing LTS and Python Skin

| Parameter | LTS | Python skin |
|---|---|---|
| Total Area Ratio (TAR) | $\dfrac{\text{Total area of dimples}}{\text{total surface area}}$ | $\dfrac{\text{Total area of fibril tips}}{\text{total surface area}}$ |
| Dimple Slenderness Ratio (DSR) | $DSR_p = \dfrac{h_p}{\Phi_p}$ | $DSR_s = \dfrac{h_s}{\Phi_s} \approx \dfrac{R_t}{\Phi_s}$ |
| Surface Aspect Ratio (SAR) | $SAR_p = \dfrac{\lambda_p + \Phi_p}{h_p}$ | $SAR_s = \dfrac{\lambda_s + \Phi_s}{h_s}$ |

## 5. Materials and Methods

All observations reported herein pertain to shed skin obtained from five male Ball pythons (Python regius). *Skin shedding in snakes occurs naturally; as such no animals were injured in obtaining the examined skins.* All the received shed skin was initially soaked in distilled water kept at room temperature for two hours to unfold. Following soaking, the skin was dried using compressed air and stored in sealed plastic bags.

To determine the parameters needed for comparison we identified thirty scales on the shed skin. The chosen scales cover the distance between the first ventral scale and the scale immediately preceding the anal opening of the reptile. The interspacing between the scales is 25 mm (centerline-to-centerline). For each of the chosen scales, we measured the lengths of the longitudinal chord (BB) and the horizontal chord (AA). The measurements then were used to calculate the aspect ratio and the area for each of the ventral scales and the fibrils.

For each chosen scale a series of five SEM pictures (at a magnification of x=10,000) at different locations within the particular scale were recorded. Analysis of these pictures yielded fibril geometric information (counts, distance between fibrils, and length of individual fibrils) along the particular locations on the AE-PE axis.

To extract data concerning the topography of the skin, we selected several swatches of skin from each of the thirty locations (1500 μm by 1500 μm) for examination using White Light Interferometery (WYKO 3300 3D automated optical profiler system). Analysis of all resulting White light Interferograms, to extract the surface parameters used two software packages: Vision®v. 3.6 and Mountains® v 6.0.

## 6. Metrological features of Python skin

### 6.1 Dimensional metrology

As mentioned in section 3, the ventral side of the reptile comprises non-uniform hexagonal unit cells (scales). The layout of these cells is such that the lateral diagonal (i.e., the diagonal in the direction of the lateral axis) is considerably longer than the diagonal in the AE-PE-direction. Moreover, the orientation of the scales is such that the lateral diagonal is perpendicular to the main direction of motion (along the AE-PE-axis for rectilinear locomotion).

The variation in the length of the lateral and longitudinal diagonals of the ventral scales implies that the aspect ratio of the building block of the ventral side is greater than unity. It appears, moreover, that the aspect ratio for the ventral scales is not uniform along the AE-PE-axis. Therefore, as a point of entry to metrological characterization we determine the aspect ratio of the ventral scales and the micron-sized fibrils within. Such a step assumes importance in light of observations by other authors that increasing the aspect ratio of a hexagonal padding, above unity, contributes toward considerable reduction of friction [55].

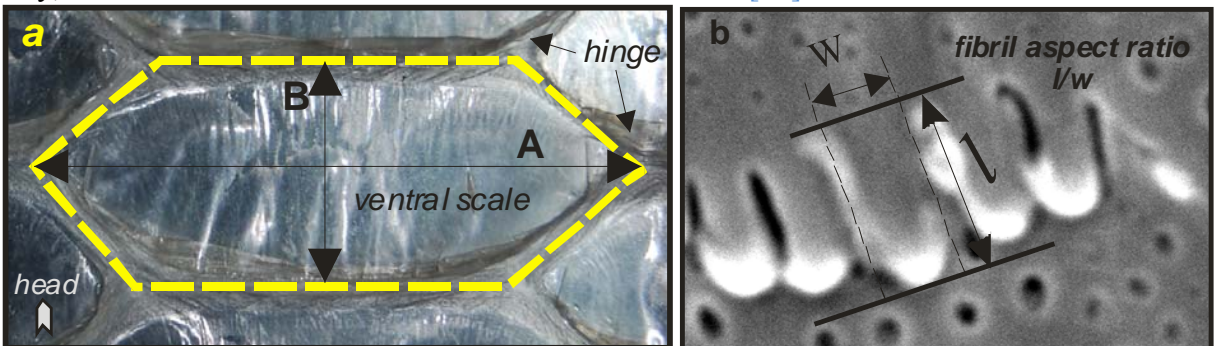

*Figure 6 Definition of the geometric parameters used for scale characterization, (a) definition of the surface area of the ventral scale, $A_{vs}$, and Ventral Scale Aspect Ratio (VSAR), (b) definition of the fibril aspect ratio.*

For a hexagonal configuration, the Ventral Scale Aspect Ratio (VSAR) may be calculated from:

$$VSAR = \frac{L_{Lateral}}{L_{(AE-PE)}} \quad (1)$$

Where, $L_{(AE-PE)}$ is the length of the chord in the direction of the AE-PE axis (line BB in figure 6-a) and $L_{lateral}$ is the length of the chord in the direction of the lateral axis (line AA in figure 6-a). Similarly, the aspect ratio of an individual fibril may be calculated from:

$$FAR = \frac{L_{fib}}{W_{fib}} \quad (2)$$

Where $L_{fib}$ is the length of the individual fibril and $W_{fib}$ is the width of the base of the fibril (see figure 6-b).

Figure (7-a) depicts the variation in the aspect ratio of the scale VSAR (blue circles) and the aspect ratio of the fibrils FAR (red circles) along the AE-PE-Axis. Dashed lines accompanying the plots represent the statistical quadratic fit for the data points. In the figure, each data point represents the average of ten measurements from different hides. The raw data does not manifest significant variation. Rather the difference between the maximum and the minimum values is small ($2.256 \leq VSAR \leq 2.606$ (refer to table 5).

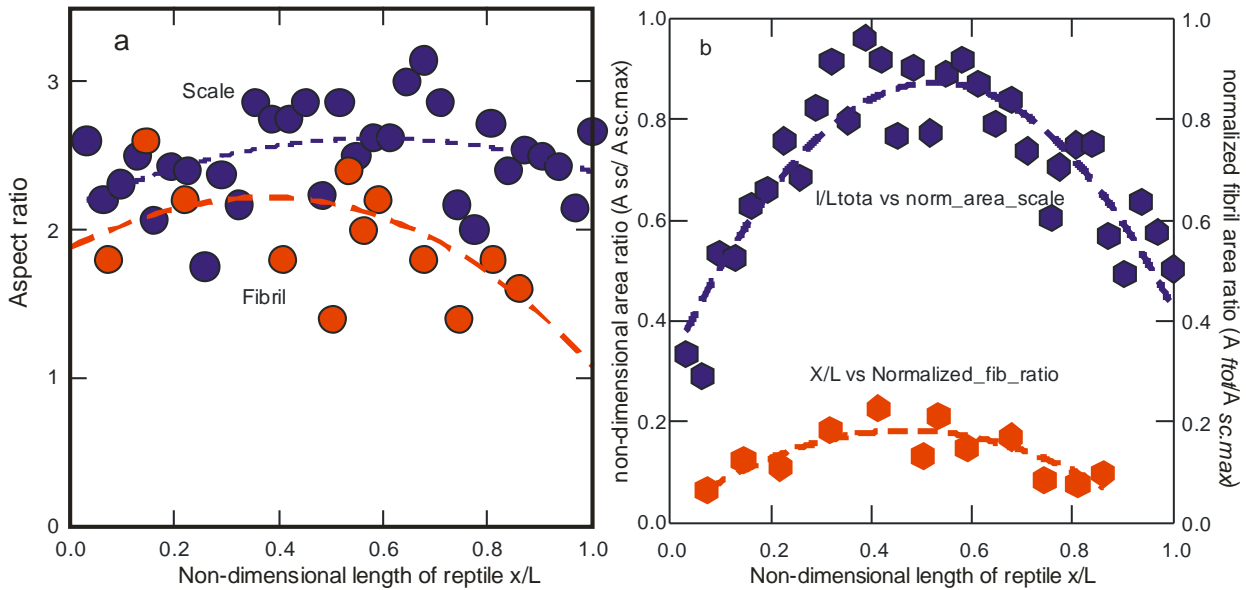

*Figure 7 Distribution of the critical geometrical attributes of the ventral scales along the AE-PE axis. Figure (7-a) distribution of the aspect ratio of the unit scale and the aspect ratio of the fibrils along the AE-PE – axis, Figure (7-b) distribution of the normalized area of the scales and the scale area ratio (area occupied by fibrils to area of scale) along the AE-PE-Axis*

Table 5 Summary variation of the key geometric attributes of the ventral scales within the leading and the trailing halves of the Python.

| Position | Front Half | | | Trailing Half | | |
|---|---|---|---|---|---|---|
| *Geometrical Parameter* | max | min | average | max | min | average |

| Area of a ventral Scale mm$^2$ | 127.25 | 66.82 | 104.75 | 120.447 | 86.282 | 103.59 |
| --- | --- | --- | --- | --- | --- | --- |
| Aspect ratio of a ventral Scale | 2.61 | 2.25 | 2.47 | 2.61 | 2.487 | 2.549 |
| Fibril density n/mm$^2$ x 10$^{-5}$ | 4.94 | 2.93 | 3.63 | 6.04 | 0.783 | 3.46 |
| Length of individual fibril μm | 1.3 | 0.7 | 1.02 | 1.3 | 0.241 | 0.825 |
| Fibril aspect ratio | 2.6 | 1.4 | 2.03 | 2.6 | 0.482 | 1.65 |
| Area of an individual fibril μm$^2$ | 0.65 | 0.35 | 0.51 | 0.65 | 0.129 | 0.41 |
| Area Ratio Fibrils/scale | 0.24 | 0.129 | 0.18 | 0.31 | 0.049 | 0.14 |

Note that the distribution of the *VSAR* is asymmetrical with respect to the AE-PE axis. The values of the *VSAR* for scales located within the trailing half of the reptile are rather higher than the values pertaining to scales located within the leading half. The average values of the *VSAR* for each half, however, are almost equal (see table 2). The maximum value for each half, however, is practically invariable, whereas the minimum value for scales in the trailing half is slightly higher than that for the leading half. The maximum overall value of the *VSAR* belongs to scales located within the region in the middle of the trunk (VSAR ≥ 3) where the highest concentration of mass takes place.

Unlike the VSAR, the FAR displays considerable variation along the AE-PE axis. In particular, it drops toward the trailing half of the body. The variation results from the change in the length of the fibrils rather than from the width of the fibril base. Referring to table 2, the maximum fibril length (1.3 μm) is the same within both halves of the body. However, the location of that maximum is different within each half. For the leading half of the body, the maximum length falls almost at the boundary of the thickest part of the trunk (which is approximately the end of the leading half). For the trailing half that maximum value falls at the beginning of that body region (almost at the MP-S section), past this location the length drops considerably. The width of fibrils, meanwhile, is almost invariable. The minimum FAR for the leading half of the reptile is almost three times the corresponding value within the trailing half. Similar to the VSAR, the distribution of the FAR is asymmetric. Fibrils located within the trailing half of the body are stout compared to those located within the leading half of the reptile.

Figure 7-b depicts the variation of the area of individual ventral scales along the AE-PE axis. The values plotted in the figure (hexagonal markers) represent the ratio between the area of the particular scale and the area of the largest ventral scale. The later was located at the middle section of the reptile. The dashed line represents a quadratic best fit. Area of scales increases toward the middle section of the trunk. Ventra scales with the largest area are located at the thickest (stockiest) region of the trunk. Thus, the largest area accommodates the heaviest cross section of the reptile. Past the middle section, area of the scales decreases until they reach their minimum value just ahead of the anal opening.

The maximum scale area within the leading half of the body is slightly larger than that within the trailing half. The minimum area, however, is smaller for the leading half compared to that within the trailing half (see table 5). Accordingly, ventral scales in the proximity of the tail region are larger than scales located at the proximity of the neck region.

For complete characterization, there is a need to find the distribution of the ratio of the area occupied by the fibrils within the particular scale. This ratio is calculated by multiplying the number of fibrils present in a particular scale by the average area of one fibril, then dividing by the total area of the scale. Repeating this process for the chosen ventral scales yields the distribution of that ratio along the AE-PE axis. This distribution is represented by the second plot within figure 7-b (red hexagonal markers and right hand side y-axis). Similar to the behaviour of the area distribution, the ratio of fibril area peaks at the middle section of the trunk. Again, this is where the mass is mostly concentrated. Beyond the middle-section, the ratio $A_{fib}/A_{scale}$ decreases. The values listed in table 5 indicate that while the average value within the leading and trailing halves of the body are invariable. The maximum and minimum values of $A_{fib}/A_{scale}$ are different with the area ratio being minimal toward the posterior end of the reptile.

As mentioned elsewhere [41], fibrils of maximum length for the python regius were located within three regions: the top and the bottom boundaries of the trunk, and the mid-section. Their lengths were found to fall within the range $1.3 < l < 1.5$ μm. The shortest fibrils meanwhile, were located within two regions: the head-neck region, and the trailing end of the load bearing volume. Their length was approximately 0.8 μm.

Figure 8 presents a plot of the internal spacing, λ, (distance between fibril rows in μm), as a function of the non-dimensional distance x/l. Particular physical location on the snakeskin may be obtained by comparing the x/l values to entries in table 1. Data plotted in the figure represent the average of five separate measurements on different regions of the same SEM picture.

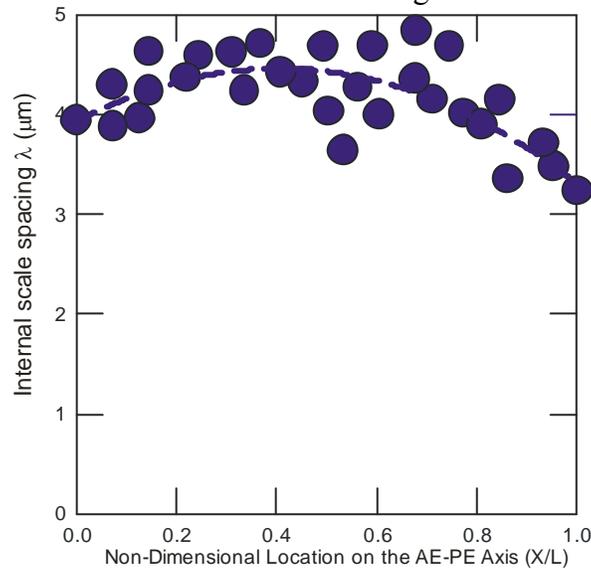

*Figure 8: Distribution of the inter-spacing between ventral fibril rows along the non-dimensional length of the reptile. (Regression line $\lambda = -2.9149x^2 + 2.3876x + 3.948$, $R^2=0.0034$)*

The distribution of the separation distance λ along the body is non-uniform. Internal spacing is larger within the trunk; the maximum is located roughly within mid-section (MP-S). The distance between fibril waves vary between 3.5μm $<\lambda<$ 4.8μm. The shortest spacing (approximately 3.5μm) is roughly located within the non-load bearing portions of the body (i.e.; the head and tail sections).

## 6.2 Topographical Metrology of Skin Surface

To determine the average profile roughness of the scales we utilized White Light Interferometery. Three scales within each of the skin pre-identified regions (i.e. TDP-S, MP-S, and BDP-S) were chosen at random. For each scale, five White Light Interferograms, WLI, were recorded. Each WLI covered a square area of about 150 μm by 150 μm.

Since the size of fibril tips, the focus of the analysis, are of the order of magnitude as small roughness, statistics of roughness profiles would be more pertinent. To extract roughness parameter information there is a need to filter form data out of the WLI. To this end, WLI for each spot on the scales were first treated to remove form data. Thereafter, we evaluated the average roughness, Ra, for small-scale profiles in the lateral and the AE-PE directions. Table 6 lists the extracted $R_a$ values in both directions (lateral and AE-PE).

Table 6 Values of the Profile Arithmetic Mean Deviation, Ra (μm), at selected regions within the skin

| Region | AE-PE-Axis | SD | Lateral-Axis | SD |
|---|---|---|---|---|
| TDP-S | 0.041 | ± 0.001 | 0.126 | ± 0.01 |
| MP-S | 0.075 | ±0.0072 | 0.06 | ±0.00813 |
| BDP-S | 0.039 | ± 0.0085 | 0.14 | ± 0.00624 |
| Average | 0.052 | ± 0.0056 | 0.1087 | ± 0.0081 |

The data show that the roughness in the AE-PE direction differs from that in the transverse direction.

## 7. Results

This section presents the results of computing the performance indicators for the skin of the python. Table 7 presents a summary of the computed results. The table includes two data sets. The first pertains to the AE-PE axis. This data set resulted from using the Ra values for profiles along the AE-PE axis. The second set meanwhile resulted from using the Ra values of profiles along the lateral axis. It is noted that the TAR value for both directions is equal. This is a result of the definition of the TAR parameter, which in essence is the percentage of the area occupied by the fibrils within the area of the scale. This is not direction dependant. Values of the DSR and the SAR manifest some difference. For each of the computed quantities we also included the average value (entries in the last raw of table 7).

Table: 7 values of the computed performance indicators along the two main axis of the skin the AE-PE and the lateral axis

| Region | AE-PE-Axis | | | | Lateral Axis | | | |
|---|---|---|---|---|---|---|---|---|
| | Ra | TAR | SAR | DSR | Ra | TAR | SAR | DSR |
| TDP | 0.041 | 0.21 | 55 | 0.1 | 0.126 | 0.21 | 43.65 | 0.13 |
| MP | 0.076 | 0.26 | 58.51 | 0.076 | 0.06 | 0.26 | 91.67 | 0.06 |
| BDP | 0.039 | 0.1 | 56.12 | 0.039 | 0.14 | 0.1 | 39.28571 | 0.14 |
| Average | 0.052 | 0.19 | 56.54 | 0.072 | 0.109 | 0.19 | 58.20 | 0.109 |

The slight difference between the computed values along the axes is rather deceptive. This is because of the sensitivity of the data and the original size of the surface features involved. To illustrate this point we introduce figure 10 (a and b). The figure is a plot of the effect of the textural anisotropy on the computed performance parameters.

Values at the abscissa represent the textural anisotropy in Ra (defined as the ratio of Ra, in the lateral direction, to that in the AE-PE direction). The quantity represented in the ordinate differs according to the figure. In figure 9-a, the ordinate depicts data for the SAR anisotropy ratio (defined as SAR in the lateral direction to the SAR value in the AE-PE direction). For figure 9-b, however, the ordinate depicts data for the DSR anisotropy ratio (defined as DSR in the lateral direction to the DSR value in the AE-PE direction). In both plots, the dashed lines represent a quadratic regression function. The symbol labeled "average" in each plot marks the average value of the particular ratio of anisotropy, which was computed, based on average values.

The DSR anisotropy ratio manifests a trend that contrasts that of SAR anisotropy ratio. In particular, the DSR ratio increases with textural anisotropy whereas SAR anisotropy drops with textural anisotropy. A question that arises given the difference in directional values is: which of the data sets to be used upon comparing the textural construction of the skin to recommended optimal values for LTS? The answer depends on the principle direction of motion of the surface under consideration. Pythons are heavy snakes. Due to their weight, they move via rectilinear locomotion where the displacement is mainly along the AE-PE axis of the reptile. As such, the use of data along the AE-PE axis in all comparisons is appropriate.

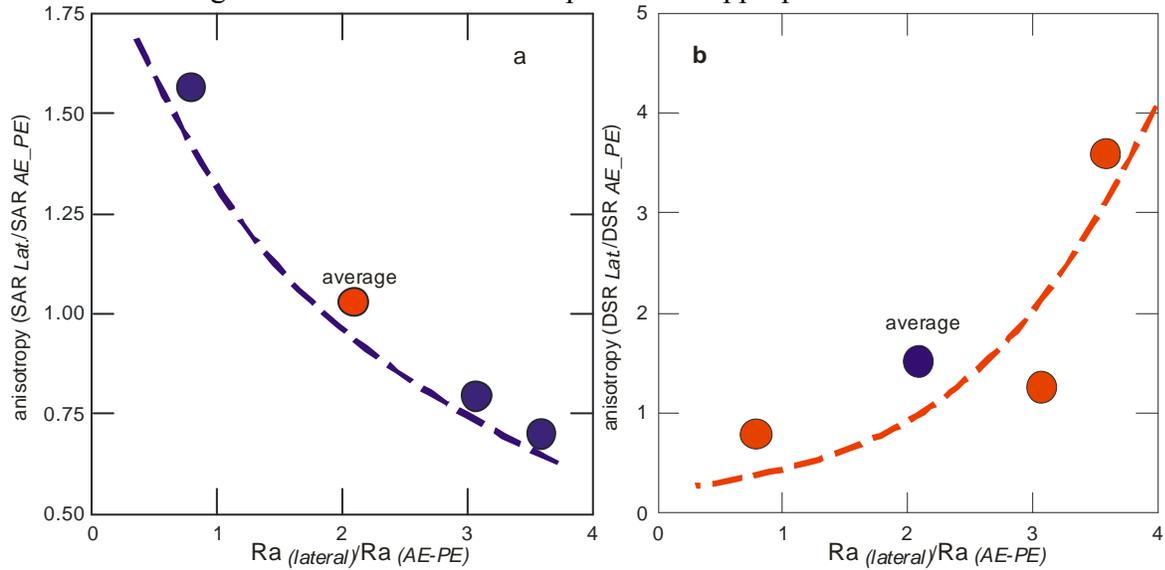

*Figure 9 Variation in the Ratio of Anisotropy for the SAR and DSR with textural anisotropy*

## 8. Discussion
### 8.1 Comparison to Laser textured surfaces

In this section, we draw a "*performance prediction*" comparison between the surface of the ventral side of the python and LTS. The scope of the comparison at this stage of the investigation is confined to comparing the numerical ranges of the performance indicators as computed from the geometry of the skin to those reported in tribology literature for LTS The comparison process entails constructing *performance indicator maps* (whenever possible) from available data. Thereafter, we superimpose the computed performance indicators for the skin on these maps. The performance indicators for the skin of the snake, as mentioned earlier, are functions of metrological parameters (both dimensional and topographical). The metrological parameters, on the other hand, differ by position and direction of profile extraction (refer to table 6). As such, performance indicators for the snakeskin are functions of the evaluated region on the body of the reptile. To generalize the comparison, we compute four variations of the skin performance indicators (and thereby we superimpose these on the constructed maps). As such, we evaluate the performance indicators for each of the pre-identified regions on the skin (TDP-S, MP-S, and BDP-S) using local data in addition to an average value for each of the comparison ratios. This average value is a function of the averaged metrological values. In this manner we identify which region, if any, is more suitable for in depth analysis that leads to surface-mimicry (i.e., simulations and experimental friction studies for surface replicas). The starting point of the devised procedure is to compare the shed skin to the experimental parameters of Kovalechenko et al [56] who performed extensive experiments to rate the tribological performance of LTS.

Kovalchenco and co-workers used a speed range of 0.15-0.75 m/s and nominal contact pressures that ranged from 0.16 to 1.6 MPa. They also used two lubrication oils with different viscosities (54.8 and 124.7 cSt at 40 $^{o}$ C). Table 3 presents a summary of the parameters describing the surfaces used in these and the equivalent parameters for the snakeskin.

Table 8 Surface designation and parameters used in the experiments of Kovalchenko et al [56]

| Surface | Description |
|---|---|
| S-3 | Standard LTS (SLTS) |
| S-4 | Higher Dimple Density (HDD) |
| S-5 | Standard unlapped (SU) |
| S-6 | Lower dimple density (LDD) |
| P-AV | Python skin based on averaged metrological quantities |

| Parameter | S-3 SLTS | S-4 HDD | S-5 SU | S-6 LDD | Python (average) |
|---|---|---|---|---|---|
| Depth of dimples h (μm) | 5.5 | 5 | 6.5 | 4 | 0.076 |
| Surface roughness between dimples Ra (μm) | 0.03 | 0.06 | 0.07 | 0.09 | 0.05 |
| Diameter of dimples $\Phi$ (μm) | 78 | 58 | 80 | 58 | 1.0 |
| Distance between dimples $\lambda$ (μm) | 200 | 80-100 | 200 | 200 | 4.5 |
| Dimple area density | 0.12 | 0.15 | 0.12 | 0.7 | 0.1 |

Figure 10 presents a summary of the results obtained by Kovalchenko and co-workers. The figure depicts evolution of the coefficient of friction (COF) of the LTS used in the experiments. Data pertain to lubricated sliding in the presence of high viscosity motor oil. From the figure, we can distinguish the general trend of the COF. Depending on the sliding speed, two regimes take place upon sliding. The first is characterized by high friction associated with low load carrying capacity. The second regime meanwhile takes place at higher sliding speeds. Here the fluid film builds up to reach a thickness capable of establishing a hydrodynamic lubrication regime. Upon increasing the sliding speed, the thickness of the fluid film increases and the load carrying capacity increases consequently. This causes the friction to drop.

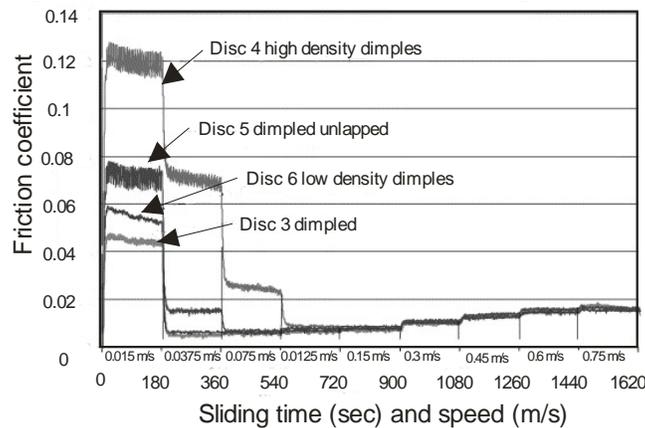

*Figure 10: Friction behavior of the surfaces used by Kovalchenko et al* [56]

The value of the COF and the rate of transition to a full hydrodynamic lubrication regime constituted the criterion used to rate the tested surfaces. Accordingly, a better performing surface, thus, would exhibit low COF, and will transit to a full hydrodynamic regime at a sliding speed lower than that of a poor performing surface. As shown in figure 10, the best performing surface is surface S-3 since it exhibits the lowest COF. Surface number S-6 exhibited the second lowest COF. As such, it was rated the second best performing surface and so on.

In principle the ranking of the surfaces depends on the ability of the lubricating oil to establish a film of sufficient thickness capable of supporting the contact loads and whence separating the rubbing surfaces. Ability of the lubricant to build a suitable film depends on the optimization of texturing manifested in optimal values of performance indicators. Naturally, not all parameters are equally influential on lubrication quality and friction reduction.

One may classify surface geometrical parameters in two categories: primary and auxiliary. Primary parameters directly influence the quality of lubrication; meanwhile auxiliary parameters play a role only when two surfaces share the same value of a primary parameter (or primary parameter set). For example, in table 6 surfaces S-3 and S-6 share the same distance between dimples, yet the COF exhibited by surface S-3 is less than that exhibited by surface S-6. This indicates that the better performance of surface S-3 relates to the difference in other parameters (dimple area density and dimple diameter for example). Similarly, surfaces S-3 and surface S-5 share the same dimple area density (0.12). Yet, surface S-3 produces a lower COF than surface S-5. So that the different values of the roughness parameter, Ra, and the diameter between the two surfaces for example should be the source of the contrast in performance. As such, complete mapping of the performance of tested surfaces in relation to their respective individual surface parameters (both primary and auxiliary) allows predicting the performance of untested surface.

To illustrate this point, suppose that we construct a Ra-TAR map for which the respective parameters of the surfaces given in table 6 constitute the entries. Suppose, further, that we plot the Ra and TAR values of a surface of unknown performance on the same map. Then the ranking of the unknown surface will depend on its location with respect to surfaces S-3 and S-6. If the new surface falls between surfaces S-3 and S-6, then it is likely to perform better than surfaces S-4 and S-5. The overall ranking of the new surface also depends on how close it is located to surface S-3 on the Ra-TAR map. Repeating such a process with maps constructed by plotting the combination of parameter pairs of the surfaces in table 3, allows ranking of the unknown surface. In what follows we apply this procedure to rank the projected tribological performance of the skin. Tables 7, 8, and 9 contain the data necessary to rank the python skin with respect to the test surfaces used by Kovalechenko.

Table 9 value of parameters used in comparing LTS and Python Skin

| Surface | DSR | TAR | SAR |
|---|---|---|---|
| S-3 | 0.0705 | 0.12 | 50.5 |
| S-4 | 0.0862 | 0.15 | 31.6 |
| S-5 | 0.0812 | 0.12 | 43 |
| S-6 | 0.0689 | 0.07 | 64.5 |
| Python (average) | 0.076 | 0.1753 | 56.54 |
| TDP-S | 0.1 | 0.214 | 55 |
| MP-S | 0.07 | 0.26 | 58.51 |
| BDP-S | 0.039 | 0.1 | 56.122 |

Figure 11 (a-d) depicts four maps: Ra versus DSR (figure 11-a), DSR versus SAR (figure 11-b), Ra versus Total Area Ratio TAR (figure 11-c), and SAR versus TAR (figure 11-d). Figure 11-a shows that the Ra parameter is not a principal performance influence parameter. The DSR value is what primarily distinguishes between the better performing surfaces and those with lower performance. Surfaces S-3 and surface S-6 fall approximately on the same DSR coordinate (the DSR values for both surfaces are so close-refer to table 8). Being almost equal, the Ra value would be the value that determines the quality of tribological performance. In this sense, the Ra value is an auxiliary influence parameter. The general trend exhibited by the data is that the performance is inversely proportional to the Ra value when the DSR values are equal. Consequently, if the snakeskin is to exhibit better performance it should be located almost on the same coordinate as S-3 and S-6. This is indeed the case (observe the location of the hexagonal symbol in the figure which denotes the average value for the python). Moreover, the python skin falls closer to S-3 than to S-6 (due to higher Ra than that of S-6) whence it is projected to perform better than S-6. To this end, a surface that mimics the python surface should fall second to S-3 under the same test conditions.

Following the same criterion, the region, on the skin, that closely matches the better surface performance band is that for the mid-region MP-S. Compared to the average value, P-AV, however, the performance of a surface mimicked after the texture of the MP-S is predicted to be less efficient than S-3. Surfaces constructed after the textural features of the TDP-S region and the BDP-S regions fall out of the band of acceptable performance.

The DSR in itself, however, may act as an initial filter for surfaces when mapped against the SAR, figure 11-b. Here, a narrow band of SAR values designates the best performing surfaces. Surfaces S-4 and S-5, for which the tribological performance is of lesser quality, fall at the higher end of the DSR axis. A linear fit of the data (the straight line shown in the figure) implies that the quality of performance is directly proportional to the SAR (the higher the SAR the better the performance).

Interestingly, the snakeskin (pentagon symbol) is closer to S-3 and falls directly on the linear fit. Similar to the trend reflected in figure 11-a, among all regions within the skin texture within the MP-S region is predicted to offer the best performance.

Upon considering the variation in the Total Area Ratio (TAR) with the Ra value (figure 11-c) it is noticed that a lower Ra value is a precondition for better performance for equal TAR values. Surfaces S-3 and S-5 share the same TAR value (0.12), yet S-3 exhibited better performance in the experiments. The skin of the Python, square symbol, falls close to S-3 (star symbol). Note that the Ra value for the Python is smaller than that of S-6. When, however, the SAR is plotted against the TAR (figure 13-d), most of the snake textures fall out of the performance band except for the BDP-S region, which falls almost mid-distance between S-3 and S-6. The BDP-S region, moreover, seems to be a good fit to the linear regression of the data.

In all, in the four plots the python skin always fell between the two highly ranked surfaces (S-3 and S-6). The python surface (whether localized or average), additionally, was mostly located closer to S-3 than S-6. Very good agreement with the linear fit of the data presented in figure 11-b was observed. Such an agreement implies feasibility of more in-depth investigation of LST that mimics the construction and geometry of the python surface.

## 8.2 Comparison to recommended values

Shinkarenko and co-investigators [57] provided optimal ranges for key performance indicator parameters. They deduced the range of an optimal DSR to fall within the interval 0.06<DSR< 0.08. In comparison, the equivalent value for the Python, see table 7, is 0.076. Further, they recommended a TAR value within the bounds 0.05 <TAR<0.5 with 0.2 being predicted as the optimal recommended value. The analogous value for the Python, again from table 7, is 0.175.

Costa and Hutchings [49] extensively investigated the influence of surface topography on lubricant film thickness. They measured film thickness using a capacitance technique in reciprocating sliding of patterned plane steel surfaces. The study employed circular patterns (pockets), grooves, and chevrons. They also varied the sliding orientation relative to the texture. The results of Costa and Hutchings concluded that a TAR of 0.11 seems to achieve the maximum film thickness for circular pockets. It is interesting to note that the equivalent ratio for the Python (0.1) fits closely to that recommended value. The same authors also found that a sample with a TAR ratio of 0.06 produced maximum film thickness, when the DSR was about 0.07. Consequently, they rationalized that such a value is optimal for texturing. Again, in comparison, the average value for the Python (0.076) falls very close.

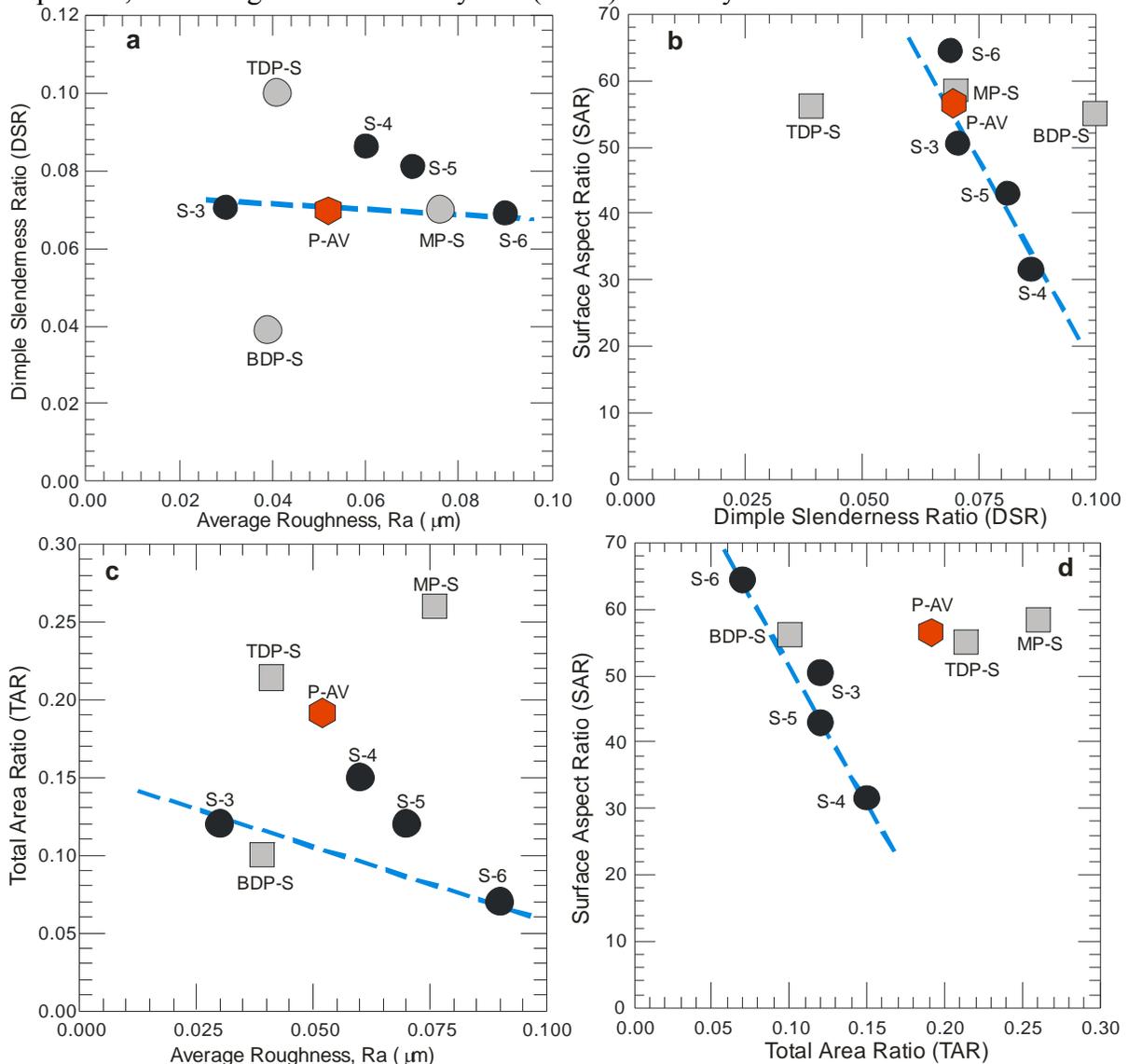

*Figure 11 Comparison maps between snake skin and Laser Textured Surfaces, a- Ra versus DSR b-DSR versus SAR, c- Ra versus Total Area Ratio TAR, d-SAR versus TAR.*

Table 10 presents a collective comparison between the performance indicator values computed for the snakeskin and those recommended by several authors. To visualize the extent of agreement between the skin values and the recommended values data in table 9 we present figure 12. The figure depicts plots for three performance indicators: TAR, DSR, and SAR (note that the figure plots the reciprocal of the later quantity for compatibility with literature values). The plot indicates that the values for the python skin are in good agreement with the recommended values.

Table 10 Summary of recommended performance indicators extracted from open literature

| TAR | DSR | SAR | Reference |
|---|---|---|---|
|  | 0.06-0.08 | 0.05-0.5 | Shinkarenko et al [57] |
| 0.025-0.1 | 0.07 | 0.11 | Costa and Hutchins [49] |
|  | 0.05 |  | Dongsheng et al [50] |
|  |  | 0.05-0.2 | Ronen et al [8] |
| 0.25 | 0.02-0.25 |  | Halperin et al [58] |
| 0.05 | 0.01-0.02 |  | Wang et al [59] |
| 0.03-0.12 |  |  | Yu and Zhou [60] |
| 0.6 |  |  | Gonzalo et al [61] |
| 0.075-0.2 | 0.03-0.08 |  | Galda et al [62] |
| 0.085 |  |  | Hu, Hu [63] |
| 0.2-0.4 | 0.03-0.1 |  | Yin et al [64] |
| 0.1753 | 0.076 | 0.017 | Python (average) |
| 0.214 | 0.1 | 0.0182 | TDP-S |
| 0.26 | 0.07 | 0.017 | MP-S |
| 0.1 | 0.039 | 0.018 | BDP-S |

The extracted data were obtained under different conditions (different materials, different shapes of textural elements, different lubricants and lubrication conditions, etc.,). Despite these differences, it is remarkable that most of the recommended values converge to a relatively narrow band of optimal values. More remarkable is the fit of the computed snakeskin values within this narrow band, which points at the optimal construction of the topographical features of the reptile. Such a finding encourages additional in-depth studies of surface design of other snake species.

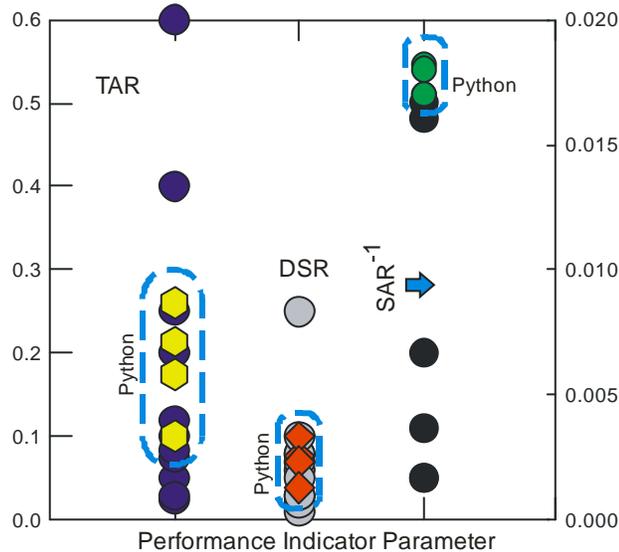

*Figure 12 Summary plot to compare the performance indicators recommended in literature and those computed for the ventral skin of the Python.*

Mimicry of reptilian surfaces does not suffice by itself for constructing a surface of optimal tribological quality. Rather it is the deduction of design rules, combined to observing reptilian dimensional proportionality that contributes to successful texturing. Observing the layout of dimples on many engineering surfaces reveals an array arrangement in which dimple arrangement is equispaced and is in patterns of parallel rows and columns. Equal spacing is hardly observed, if at all existent, in snake ventral scales (which are the principal frictional frontier in snakes). Rather, a wave arrangement is present in all sliding directions. An additional observation is the orientation of the fibrils along the body.

The orientation of the fibrils along the AE-PE axis always points toward the posterior end. The fibrils are not necessarily parallel to the AE-PE axis. This orientation contrasts the orientation of the hexagonal ventral scales (in which the lateral diagonal is always perpendicular to the AE-PE axis). Combined to the waveform distribution of spacing, the orientation features of the fibrils enhance both the friction and wear performance of the skin. Yuan et al [71] performed exhaustive studies on the effect of textural orientation on friction and wear behavior of surfaces. The authors used several LTSs in lubricated sliding. They also varied the contact pressures and the pressure of lubricant supply. Yuan and coauthors concluded that the orientation of the micro-textural elements has a strong influence on friction performance of sliding surfaces. For relatively low contact pressure small sized grooves (order of few microns), that are oriented perpendicular to the sliding direction, reduce friction by about 40% compared to un-textured surfaces. Under high contact pressure, bigger grooves (order of 20 μm), but *parallel* to the sliding direction, yield better friction performance and reduced contact stresses. The best performance reported in the work of Yaun pertained to textural elements oriented at an angle to the principal direction of motion (regardless of size). This orientation resulted in reducing the friction by a factor of two. It is interesting to compare the findings of Yaun and coworkers to the texturing pattern observed on the ventral side of the Python (figures 3-b, d, and f and figure 4). The distribution of the fibrils, as revealed from the SEM photographs, is a wave of varying amplitude and period. More important, however, most of the fibrils make an angle with the AE-PE axis, which matches the findings of Yuan, and point at the possibility that ventral structure of the python is optimal for tribological function. Naturally, to generalize such a statement there is a need for extensive work to characterize the local friction response of multiple species and compare it to the ventral structure. Such an effort is currently undertaken in our laboratory.

**Conclusions**

In this work, we presented a comparison between the structure and geometrical metrology of the ventral skin of Python regius and industrial Laser Textured surfaces.

Dimensional and metrological performance parameters, influential to efficient tribological function of textured surfaces, were found to be optimal in the case of the reptile. This, points at the feasibility of investigating more bio-analogues for developing a standard of performance that indexes the various Laser Textures currently in application.

Several fundamental differences in both geometry and construction of the compared surfaces were noted. In particular, within the reptilian surface, arrangement of surface motifs is a-periodic, asymmetric and follows a wave pattern. This allows the surface to condition its tribological response upon sliding. Such an arrangement is not practiced in manufactured surfaces where dimples are positioned in matrices formed of perpendicular rows and columns.

The ability of reptilian surfaces to self-adapt in response to changes in sliding conditions seems to originate from the holistic optimization that considers surface form, topography and morphology. This consideration may form a foundation for a generation of deterministic surface textures in fabricated surfaces.